\documentclass{aa}
\usepackage{graphicx,natbib}

\newcommand{\micron}{\ensuremath{\mu\mbox{m}}}
\bibpunct{(}{)}{;}{a}{}{,}
\begin{document}
\thesaurus{03(03.09.2)}
\title{Integrated optics for astronomical interferometry}
\subtitle{I. Concept and astronomical applications} 
\author{Fabien Malbet\inst{1}
  \and Pierre Kern\inst{1} 
  \and Isabelle Schanen-Duport\inst{2} 
  \and Jean-Philippe Berger\inst{1}
  \and Karine Rousselet-Perraut\inst{1}
  \and Pierre Benech\inst{2}}
\institute{Laboratoire d'Astrophysique UMR CNRS/UJF 5571, Observatoire de
  Grenoble, BP 53, F-38041 Grenoble cedex 9, France
  \and
  Laboratoire d'\'Electromagn\'etisme Microondes et Opto\'electronique UMR
  CNRS/INPG/UJF 5530, BP 257, F-38016 Grenoble cedex 1, France}
\offprints{F. Malbet} 
\mail{malbet@obs.ujf-grenoble.fr} 
\date{Received February 25; Accepted May 12, 1999 \hfill
  \textit{Astron.\ Astrophys.\ Suppl.\ Ser.\ } \textbf{138}, 1--10 (1999)}  
\authorrunning{F. Malbet et al.}  
\titlerunning{\small Integrated optics for interferometry. I.
  Concept and applications} 
\maketitle

\begin{abstract}
  We propose a new instrumental concept for long-baseline optical
  single-mode interferometry using integrated optics which were developed
  for telecommunication. Visible and infrared multi-aperture interferometry
  requires many optical functions (spatial filtering, beam combination,
  photometric calibration, polarization control) to detect astronomical
  signals at very high angular resolution. Since the 80's, integrated
  optics on planar substrate have become available for telecommunication
  applications with multiple optical functions like power dividing,
  coupling, multiplexing, etc. We present the concept of an optical /
  infrared interferometric instrument based on this new technology. The
  main advantage is to provide an interferometric combination unit on a
  single optical chip. Integrated optics are compact, provide stability,
  low sensitivity to external constrains like temperature, pressure or
  mechanical stresses, no optical alignment except for coupling, simplicity
  and intrinsic polarization control. The integrated optics devices are
  inexpensive compared to devices that have the same functionalities in
  bulk optics. We think integrated optics will fundamentally change
  single-mode interferometry. Integrated optics devices are in particular
  well-suited for interferometric combination of numerous beams to achieve
  aperture synthesis imaging or for space-based interferometers where
  stability and a minimum of optical alignments are wished.
  \keywords{Instrumentation: interferometers}
\end{abstract}

\section{Introduction}

Optical long baseline interferometry is one of the upcoming techniques that
will undoubtedly provide compelling, high angular resolution observations
in optical astronomy. The first attempt to use interferometry in astronomy
was proposed by \citet{Fiz68} and achieved by \citet{Ste74} on a single
telescope with a pupil mask. \citet{Mic21} first succeeded in measuring
stellar diameters, but their interferometer was not sensitive enough to
enlarge their investigation. Interferometry is a rather complex technique
which needs extreme accuracies directly proportional to the foreseen
spatial resolution: 1 milliarcsecond on the sky translates to 0.5\mbox{
  $\mu$m} in optical delay on a 100-m baseline. That is why modern direct
interferometry started only in the 70's with \citet{Lab75} who produced
stellar interference with 2 separated telescopes.  Also interferometric
experiments require very low noise detectors which became available only
recently. In addition, the atmosphere makes the work even more difficult and
dramatically limits the sensitivity of ground-based interferometers.
Space-based interferometric missions are therefore being prepared, like the
NASA \emph{Space Interferometric Mission (SIM)} or the interferometry
corner stone in the ESA Horizon 2000+ program: \emph{Infrared Spatial
  Interferometer, (IRSI)}.

Long baseline interferometry is based on the combination of several stellar
beams collected from different apertures and is aimed to either aperture
synthesis imaging \citep{Len84a} or astrometry \citep{Shao77}. A number of
interferometers are currently working with only two apertures: SUSI
\citep{Dav94}, GI2T \citep{Mou94}, IOTA \citep{Car94}, PTI \citep{Col94}.
COAST \citep{Bal96} and NPOI \citep{Ben97} have started to perform optical
aperture synthesis with three apertures by using phase closure techniques.
The increase in the number of apertures is one of the major feature of new
generation interferometers, like CHARA with up to 7 apertures \citep{McA94} or
NPOI with 5 siderostats \citep{Whi94}. We are on the verge of new
breakthroughs with the construction of giant interferometers like the VLTI
(Very Large Telescope Interferometer) by the European community which will
use four 8-m unit telescopes and three 1.8-m auxiliary telescopes
\citep{Mar98}, or the Keck Interferometer \citep{Col98} which will
have two 10-m telescopes and four 1.5-m outriggers. They will both
achieve high sensitivity thanks to their large apertures and allow the
combination of more than three input beams.

We propose in this article a new technology for beam combination that is
inherited from the telecom field and micro-sensor applications. This
technology will answer many issues related to astronomical interferometry.
The technology is called {\em integrated optics on planar substrate}, or,
in short, {\em integrated optics}. The principle is similar to that of
fiber optics since the light propagates in optical waveguides, except that
the latter propagates inside a planar substrate. In many aspects,
integrated optics can be considered like the analog of integrated circuits
in electronics. 

We describe in Sect.\ \ref{sect:intfunct} the optical functions required by an
interferometer. We present in Sect.\ \ref{sect:io} the principle of integrated
optics, the technology and the available optical functions.
Section \ref{sect:iofai} presents the concept for an interferometric
instrument made in integrated optics, and touches upon future
possibilities. Section \ref{sect:disc} discusses the different technical
and scientific issues of this new way of doing interferometry. Results with
a first component coming from micro-sensor application will be presented in
paper II \citep{Ber99}. They demonstrate the validity and feasability of
the integrated optics technology for astronomical interferometry.

\section{Description of a single-mode interferometer}
\label{sect:intfunct}

\begin{figure*}[t]
  \begin{center} \leavevmode
    \includegraphics[angle=-90,width=0.9\textwidth]{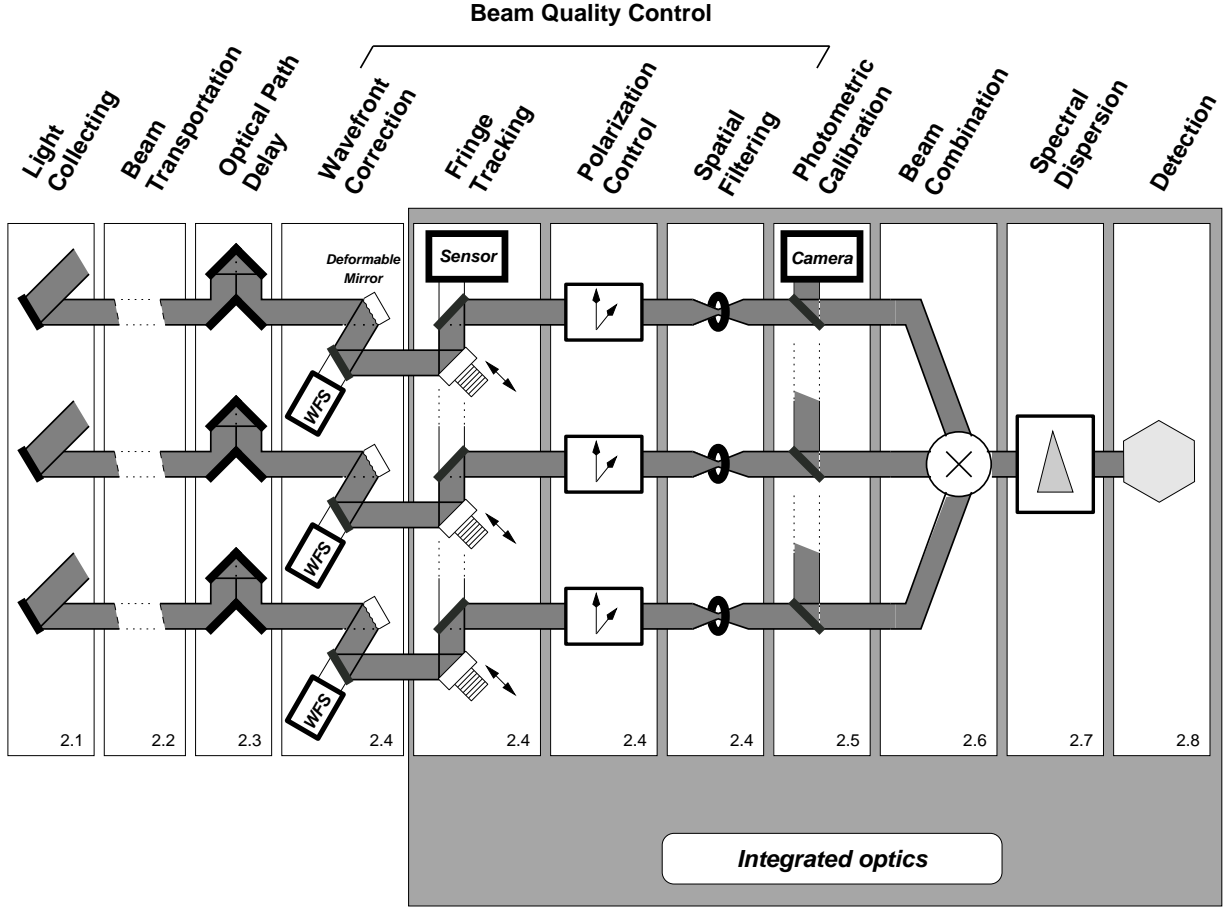}
    \caption{Functional diagram of a single-mode interferometer. The dark
    underlying box merges the interferometric functions that can be
    integrated on a single optical chip. The numbers at the right bottom refer
    to the corresponding paragraph numbers.}  
    \label{fig:functdiag}
    \end{center}
\end{figure*}

To understand where and how integrated optics can play a role in
astronomical interferometry, we review the different optical functions
present within an interferometer (see Fig.\ \ref{fig:functdiag}). This
comes after a summary of stellar interferometry principles. All
interferometers but GI2T being single-mode beam combiners (the field is
limited to the diffraction pattern of each aperture), we limit our study to
the single-mode field, the most appropriate mode for integrated optics.

A two-telescope stellar interferometer provides the measure of interference
fringes between two beams at the spatial frequency $B/\lambda$, where 
$\lambda$ is the wavelength, and $B$ the projection of the baseline vector 
$\vec{B}$ defined by the two telescopes along $\vec{s}$ the unit vector 
pointing to the source. The complex visibility of these fringes
is proportional to the Fourier transform of the object intensity
distribution (Van-Cittert Zernike theorem). Hereafter we call {\em
visibility} $V$ the modulus of the degree of coherence at the spatial
frequency $B/\lambda$ normalized to the value at the zero
frequency,
\begin{equation}
  \label{eq7}
  V =\frac{|\tilde{I}(B/\lambda)|}{|\tilde{I}(0)|},
\end{equation}
and {\em phase} $\phi$ its argument. The phase is related to the position of the 
photo-centroid of the source $\vec{s}$ by  the relation:
\begin{equation}
  \label{eqp}
  \phi =2\pi\frac{\vec{B}.\vec{s}}{\lambda}.
\end{equation}
For ground-based interferometers, the source phase is corrupted by
atmospheric turbulence. This prevents an absolute measurement of the source
phase. However it is possible to measure the difference in source phase
between two wavelengths\footnote{Phase-closure and phase-reference
  techniques also provide ways of retrieving this phase.}.

\subsection{Light collecting}

The stellar light is collected by each individual aperture. These apertures
can be either siderostats (e.g. Mark III, PTI, IOTA) or telescopes (e.g.
GI2T, VLTI, Keck Interferometer).  The coverage of the spatial frequencies
is usually done by carefully locating the apertures in order to take
advantage of the earth rotation which induces a variation of the length and
orientation of the projected baselines (super-synthesis effect). If the
structure of the object does not depend on wavelength, then observing at
different wavelengths is equivalent to observing at different spatial
frequencies. When the apertures are movable (GI2T, IOTA, SUSI), the
interferometer can cover many different baselines with different
geometrical configurations.

\subsection{Beam transportation}

The beams coming out from each telescope must be directed toward the beam
combination table. Two different techniques can achieve this transportation:
\begin{itemize}
\item Bulk optics
  
  Flat mirrors are usually used to carry the light from the single
  apertures toward the central beam combiner. Their main advantages are
  high throughput and low wavelength dependency. However they are
  sensitive to thermal and mechanical disturbance and they require many
  degrees of freedom to align the beams.
  
  Two different philosophies have been developed for transportation. 1) The
  Coud\'e trains are symmetrical to prevent differential polarization
  rotations and phase shifts. It leads to a large number of mirrors and
  thus a low throughput especially in the visible. One still get residual
  polarization effects essentially due to optical coatings differences
  which are not negligible. 2) The number of optics is reduced to a minimum
  and the large resulting polarization effects are calibrated and corrected
  inside the interferometer (Sect.\ \ref{sect:quality}).
  
\item Fiber optics
  
  \citet{Fro81} and \citet{Con84} were the first to propose fiber optics to
  connect different apertures. Major efforts have been achieved in this
  field by \citet{Sha87,Sha90,Rey94,Rey96} with silica fibers and in the
  2.2 $\mu$m range by \citet{For91, For96b} with fluoride fibers.
  
  The optical fiber throughput is very high: 100-m silica fiber has a
  throughput of $99.6\%$ at $\lambda=1.6\micron$ (0.15dB/km). In addition,
  fibers offer some flexibility since the only degrees of freedom are
  located at the entrance and output of the fiber. That is one reason why
  \citet{Tur97} have proposed optical fibers to combine the visible beams
  of CHARA. Using fibers can be significantly less expensive than bulk
  optics.
  
  The several drawbacks of using optical fibers are: chromatic dispersion
  if the optical path through the different fibers is not matched;
  mechanical and thermal sensitivity (optical fibers are also used as
  micro-sensors); and birefringence of the material. However \cite{Rey96}
  have shown that one can overcome most of these difficulties by controlling
  actively the fiber length, carefully polishing the fiber ends and by
  using polarization maintaining fibers.
\end{itemize}

\subsection{Optical path delay (OPD)}

The optical path from the beam combiner upward to the stellar source are
not identical for each beam. The interferometer must equalize the
pathlength at the micron-level accuracy. Furthermore the path lengths
change with time and the interferometer must take into account the sidereal
motion. This optical function is performed with delay lines. 

The classical solution consists in a retro-reflector based on a moving
chart \citep{Col91}. The retro-reflector can be either a cat's eye or a
corner cube. \citet{ReD94,Zha95} have proposed to stretch optical fibers to
delay the optical path. Laboratory experiments showed that this type of
delay lines can achieve more than 2 m continuous delay with 100 m silica
fibers \citep{Sim96}, and about 0.4 mm continuous delay with 3.4 m fluoride
fibers \citep{Zha95}. However in the latter case, the maximum optical path
delay is somewhat limited since the fiber length is restricted to the
maximal accepted stretch: \citet{Zha95,Mar96} proposed multi-stage delay
lines which perform short continuous delays by fiber stretching and long
delays by switching between fiber arms of different lengths. However the
differential dispersion in fibers of different length still remains a
limiting factor of this technology.

Optical path modulation using silica fibers has been implemented in the ESO
prototype fringe sensor unit \citep{Rab96}.

\subsection{Beam quality control}
\label{sect:quality}

The control of the beam quality is essential to maintain the
intrinsic contrast of the interferometer. 

\begin{itemize}
\item Wavefront correction
  
  The stellar light goes through the atmosphere where the wavefront
  is disturbed. Depending on the wavelength and the size of the turbulent
  cell ($r_0$) compared to the aperture size, the incoming wavefront is 
  corrugated and the stellar spot divided in several speckles with phase 
  differences in the focal plane.
  Single-mode interferometers select only one speckle and therefore the
  atmospheric turbulence leads to signal losses proportional to the Strehl
  ratio. Using adaptive optics to correct at least partially the incoming
  wavefront increases the total throughput of an interferometer. The
  minimum wavefront correction is the tip-tilt correction used on many
  interferometers (IOTA, SUSI, PTI,...)

\item Fringe tracking
  
  Due to the same atmospheric perturbations but at the baseline scale, the
  optical path between two apertures will rapidly vary. When requiring a
  high sensitivity like for spectral analysis, one needs to increase the
  acquisition time. The interferometric signal must be analyzed faster than
  the turbulence time scale to prevent visibility losses due to fringe
  blurring. The fringe tracker analyzes the fringe position and actively
  control a small delay line to compensate the atmospheric delay. The
  fringes are stabilized.

\item Polarization
  
  Instrumental polarization can dramatically degrade the frin\-ge visibility. 
  The main effects are differential rotations and phase shifts between the 
  polarization directions \citep{Rou96}. Even if special care is taken in 
  designing the optical path to have the most symetrical path for each beam, 
  in practice the incident angles are not exactly the same and the 
  mirrors do not have the same coatings. 

  Differential rotations can be compensated by rotator devices \citep{Rou98} 
  whereas differential phase shifts can be corrected by Babinet compensators 
  \citep{Rey93} or Lef\`evre fiber loops \citep{Lef80}.

\item Spatial filtering
  
  The incoming wavefronts propagate through a spatial filter, a geometrical
  device which selects only one coherent core of the beams. It can be
  achieved either by a micrometer-sized hole or by an optical waveguide
  like a fiber \citep{Sha88}. This principle has been applied successfully
  to the FLUOR interferometric instrument \citep{For96a}. The beams
  including atmospheric turbulence effects are then characterized by only
  two quantities, the amplitude and the phase of the outcoming electric
  field\footnote{In fact, this statement is correct only for long enough
    fibers ($>1000\lambda$ like a few centimeters) or small hole (a few
    tenths of the diffraction-limited pattern).}.  Combined with
  photometric calibration, this process leads to accurate visibilities (see
  Sect.\ \ref{sect:interf.photcal}).
\end{itemize}

\subsection{Photometric calibration}
\label{sect:interf.photcal}

The interference signal which is measured in stellar interferometry is
directly proportional to each incoming beam intensity. These intensities
fluctuates because of the atmospheric turbulence. The estimation of the
fringe contrast is improved when these intensities are monitored as
suggested by \citet{Con84} and validated by \citet{For96a}. Photometric
calibration combined with spatial filtering leads to visibility accuracies
better than 0.3\% \citep{For96b}.

\subsection{Beam combination}
\label{sect:bc}

\begin{figure}[t]
  \begin{center}
    \leavevmode
    \includegraphics[angle=-90,width=0.95\columnwidth]{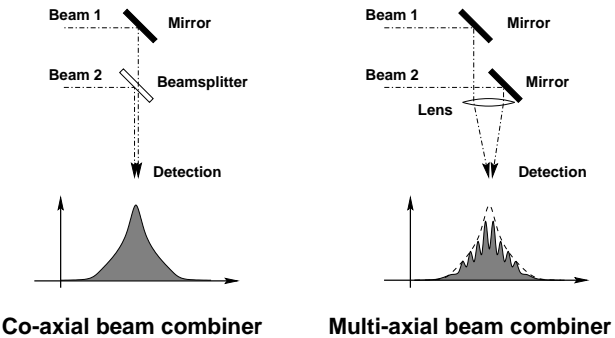}
    \caption{The different types of beam combination in classical optics
    with the profile of the output intensity.}
    \label{fig:recomb_bulk}
  \end{center}
\end{figure}

\citet{Mar92} have classified the different types of beam
combinations. In the single-mode case, there are two types of
beam combination:
\begin{itemize}
\item co-axial combination, when the beams seems to propagate from the same
  direction as in the Michelson laboratory experiment (left part of Fig.\
\ref{fig:recomb_bulk}); 
\item multi-axial combination, when the beams seems to propagate from different
  directions in the Young's double slit experiment (right part of Fig.\
\ref{fig:recomb_bulk});
\end{itemize}

In bulk optics, the co-axial combination is performed with a beam splitter
whereas the multi-axial combination is done by focusing the different beams
on the same spot. In the case of multi-axial combination, the differential
tilt between the beams produces fringes on the point spread function. The
co-axial combination can be regarded as a particular case of the multi-axial
mode where all the beams are superposed without tilts: the fringes
disappears and the amplitude of the resulting spot depends on the phase
difference between the two beams.

The fringe encoding is achieved, in the co-axial case, by modulating the
optical path difference between the two beams which results in an intensity
modulation, or, in the multi-axial case, by sampling the spatial fringes
with a detector matrix\footnote{If OPD modulation is used with the
  multi-axial combination, then the fringes appears to move underneath the
  fringe envelope.}. Usually if the fringes are coded in one direction, the
other direction is compressed to reduce the number of pixels.

\subsection{Spectral information}
\label{sect:spect}

This function is not always implemented in existing instruments, although
it is useful for two objectives: to estimate the physical parameters of the
source (temperature, kinematics,...), and, to determine the position of the
central fringe at zero OPD. The distance between the fringes being directly
proportional to the wavelength, one can derotate the fringe phase like in
Mark III and PTI \citep{Shao88} or to measure the group-delay like in GI2T
\citep{Koe96}.  

The spectral information can be obtained by dispersing the fringes with
a dispersive element (GI2T, PTI). \citet{MRi88} also suggested to apply the
concept of Fourier transform spectrography to interferometry by performing
double Fou\-rier transform interferometry.

\subsection{Detection}

In the visible, the detectors are either CCDs or photon-counting cameras.
In the infrared, for long mono-pixel InSb detectors have been used,
but with the availability of array detectors with low read-out noise,
interferometers started to use detector matrices.

\section{Integrated optics on planar substrate}
\label{sect:io}

The concept of integrated optics was born in the 70's with the development
of optical communications by guided waves. A major problem of transmission
by optical fibers was the signal attenuations due to propagation and the
need for repeaters to reformat and amplify the optical signals after long
distances.  The solution offered by classical optics was unsatisfactory and
\cite{Mil69} suggested an integrated, all-optical
component on a single chip, with optical waveguides to connect them.

\subsection{Principle of guided optics}

\begin{figure}[t]
  \begin{center}
    \leavevmode
    \includegraphics[angle=-90,width=0.6\columnwidth]{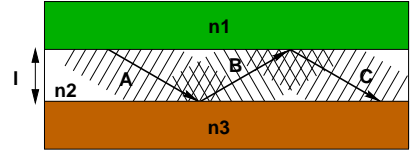}
    \caption{Principle of optical guidance (see text for details).}
    \label{fig:pringuide}
  \end{center}
\end{figure}

For sake of simplicity, we first consider the wave propagation of a
collimated incident beam into a planar waveguide. This particular structure
is formed of three step-index infinite planar layers (see Fig.\ 
\ref{fig:pringuide}). Light can be observed at the structure output
provided that total reflection occurs at each interface and constructive
interferences occur between two successive reflected wavefronts (A and C in
the figure). The first condition implies that a high-index layer is
sandwiched between two low-index layers and gives the range of acceptable
incident angle. The second condition translates into a phase difference
between the wavefronts A and C multiple of 2$\pi$. Therefore the range
of acceptable incident angles is no longer continuous but discrete. A
single-mode waveguide is a guide which can propagate only the direction
parallel to the waveguide. The core layer thickness ranges between
$\lambda/2$ and $10\lambda$ depending on the index difference. Multimode
guide propagates beams coming from different directions.

In practice, one needs the full electromagnetic field theory to compute the
beam propagation inside the waveguide. The continuity relations of the
electromagnetic fields at each interface lead to the equations of
propagation of guided modes \citep{Jeu90}.  Depending on the wavelength and
the guide thickness ($l$ in the Fig.\ \ref{fig:pringuide}), these equations
have either no solution (structure under the cutoff frequency), either only
one solution (single-mode structure) or several ones (multi-mode
structure). The number of solutions also depends on the difference of
refractive index between the various layers of the structure. The larger
the index difference are, the better the modes are confined. These
equations also allow to estimate the energy distribution profile which can
be approximated, to first order, by a Gaussian function.  The major part of
the energy lies in the channel, but evanescent waves can interact with
evanescent waves coming from other close waveguides (see the directional
coupler in Sect.\ \ref{sect:io.funct}).

In interferometry, multi-mode guided structures cannot be used since there
exist optical path differences between the various modes. In the
following, only single-mode waveguides are considered.

\subsection{Current technologies}
\subsubsection{Ion exchange}

\begin{figure}[t]
  \begin{center}
    \leavevmode 
    \includegraphics[angle=-90,width=0.95\columnwidth]{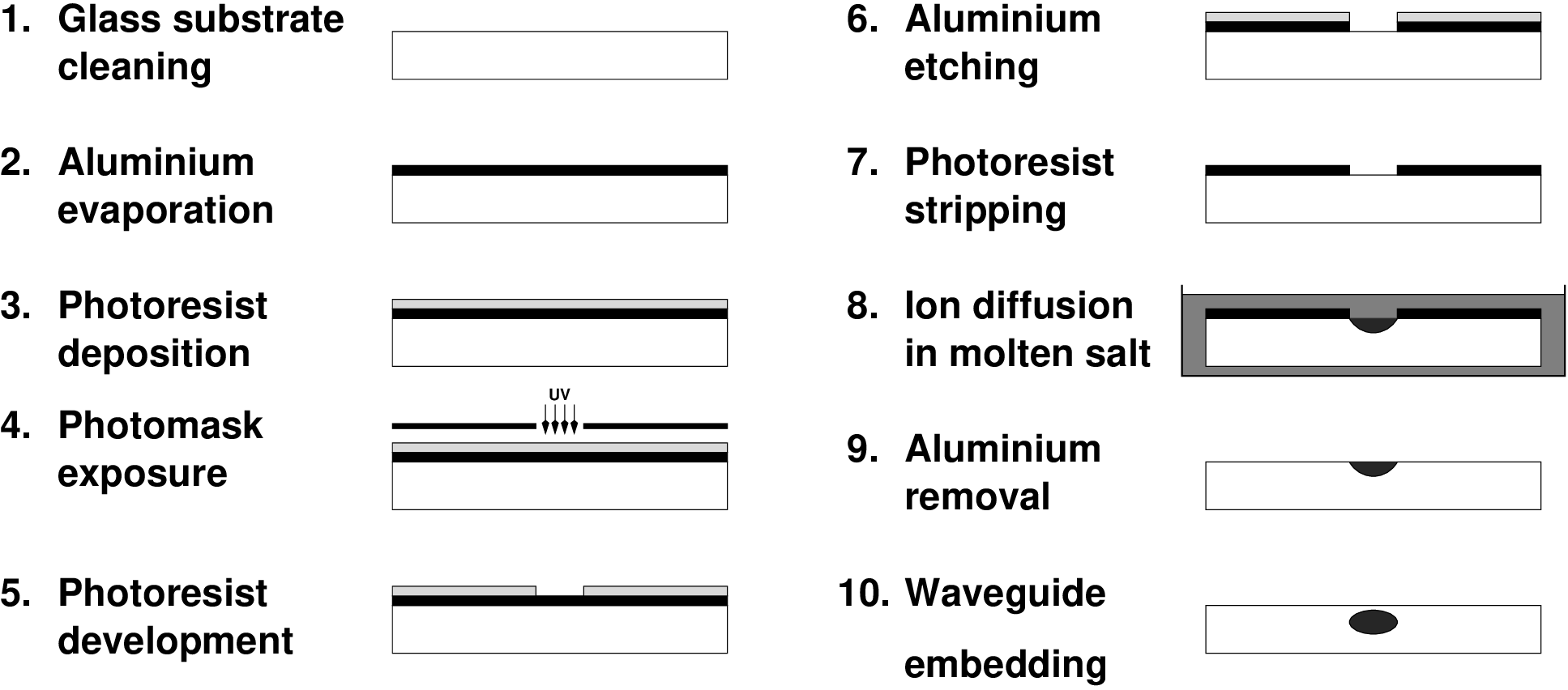}
    \caption{Waveguide manufacture by ion exchange technique \citep{Sch96}.}
    \label{fig:photolitho}
  \end{center}
\end{figure}

A first method to build integrated guides on planar substrate is based on
glass ion exchange \citep{Ram88,Ros89}: the $\mbox{Na}^+$ ions of a glass
substrate are exchanged by diffusion process with ions ($\mbox{K}^+$,
$\mbox{Tl}^+$ or $\mbox{Ag}^+$) of molten salts. The local modification of
the glass chemical composition increases the refractive index at the glass
surface. A three-layer structure (air / ions / glass) is created and the
light is vertically confined. By standard photo-masking techniques (see
Fig.\ \ref{fig:photolitho}), the ion exchange can be limited to a compact
area and create a channel waveguide. Since ion exchange only occurs at the
surface of the glass, the last step of the process consists in embedding
the guide, either by forcing the ions to migrate with an electric field or
by depositing a silica layer on the waveguide. We obtain a component which
guides the light like an optical fiber, the ion exchange area being the
core and the glass substrate\footnote{with or without the added silica
layer} being the cladding.  According to the ions of the molten salt, the
refractive index difference can vary between 0.009 and 0.1 (see
Table~\ref{tab:ions}).This technology provides various components for
telecom and metrology applications.

\begin{table}[t]
  \caption{Ions characteristics in ion exchange technology.}
  \label{tab:ions}
  \smallskip
  \begin{center}
    \leavevmode
    \begin{tabular}{llll}
      Ions&$\Delta n$&Comments\\
      \hline
      Li$^+$&0.02& High tensile stresses\\
      K$^+$&0.009& Compressive stresses\\
      Rb$^+$&0.01& High price\\
      Cs$^+$&0.04& Slow diffusion\\
      Tl$^+$&0.10& Attention for safety\\
      Ag$^+$&0.10& Low thermal stability\\
      \hline
    \end{tabular}
\end{center}
\end{table}

\subsubsection{Etching technologies}

\begin{figure}[t]
  \begin{center}
    \leavevmode
    \includegraphics[angle=-90,width=0.95\columnwidth]{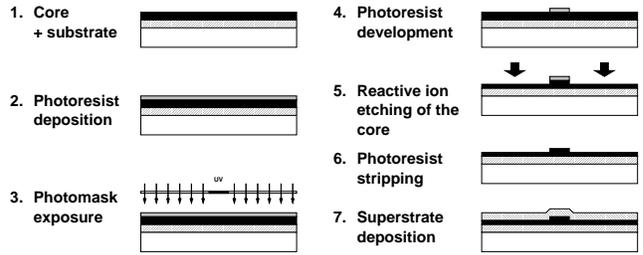}
    \caption{Waveguide manufacture by etching technique \citep{Mot96}.}
    \label{fig:silicon}
  \end{center}
\end{figure}

Another method consists of etching layers of silicon of various indices
\citep{Mot96}. These layers can be either phosphorus-doped silica or
silicon-nitride.  Both technologies can create channels by etching
layers of material, where light is confined like in an optical fiber (see
Fig.\ \ref{fig:silicon}). The channel geometry is defined by
standard photo-masking techniques. According to the fabrication process,
$\Delta n$ can be either high (0.5) for very small sensors, or very low
(between 0.003 and 0.015) for a high coupling efficiency with optical
fibers.  These technologies usually provide components for various
industrial applications (gyroscopes, Fabry-P\'erot cavities or
interferometric displacement sensors).

\subsubsection{Polymers}

Single mode waveguides made by direct UV light inscription onto polymers
are in progress. Such a technology is still in development and the
components present usually high propagation losses \citep{Stro98}.

\subsection{Available functions with integrated optics}
\label{sect:io.funct}

\begin{figure}[t]
  \begin{center}
    \leavevmode
    \includegraphics[angle=-90,width=0.6\columnwidth]{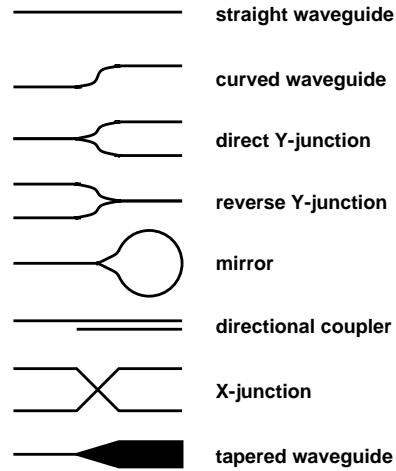}
    \caption{Available elementary integrated optics functions.}
    \label{fig:fonction_io}
  \end{center}
\end{figure}

The first two technologies provide many standard
functions for wavelengths ranging between 0.5 $\micron$ and 1.5 $\micron$ 
(standard telecom bands). Several examples are presented (see 
Fig.\ \ref{fig:fonction_io}):
\begin{enumerate}
\item The {\bf straight waveguide} is the simplest component.
\item The {\bf curved waveguide} allows some flexibility to reduce
  the size of integrated optics components. Its characteristics depend on
  the radius of curvature.
\item The {\bf direct Y-junction} acts as an achromatic 50/50 power divider.
\item The {\bf reverse Y-junction} is an elementary beam combiner similar to
  a beam-splitter whose only one output is accessible\footnote{The flux is
  lost if the incident beams are in phase opposition.}.
\item The {\bf mirror} is an Y-junction coupled with curved wave\-guides 
  creating a
  loop. A straight transition between the Y-junction and the loop ensures a
  symmetrical distribution. The modes propagating through the loop in
  opposite directions interfere and then light goes back in the input
  straight waveguide.
\item The {\bf directional coupler} consists in two close waveguides.  
  According
  to their proximity and the length of the interaction area, modes can be
  transfered between them and a power divider can be realized. The power
  ratio obviously depends on the distance between the two guides, the
  length of the interaction area and the wavelength.
\item The characteristics of the {\bf X-crossing} depend on the intersection
  angle.  For high angles (e.g. larger than 10 degrees), the two waveguides
  do not interact: the crosstalk is negligible. For smaller angles, a part
  of power is exchanged between the two arms of the components.
\item The {\bf taper} is a smooth transition section between a single-mode
  straight waveguide and a multi-mode one. It allows light to propagate in
  the fundamental mode of the multimode output waveguide. The output beam
  is thus collimated.
\end{enumerate}

\section{A coin-size complete interferometer}
\label{sect:iofai}

Many functions required by interferometry (see Fig.\ ~\ref{fig:functdiag})
can be implemented on a single integrated optics component made from a tiny
glass plate. Based on the listed available functions, one can design a beam
combiner for a multi-telescope interferometer.

\subsection{Beam combination}
\label{sect:interf.bc}

\begin{figure}[t]
  \begin{center}
    \leavevmode
    \includegraphics[angle=-90,width=0.95\columnwidth]{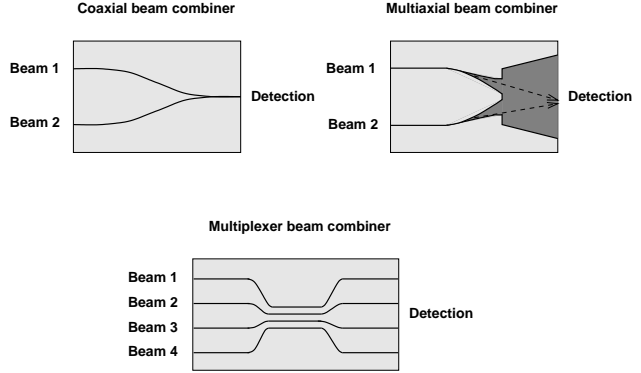}
    \caption{Different types of integrated optics beam combiners.}
    \label{fig:recomb}
  \end{center}
\end{figure}

Fig.\ \ref{fig:recomb} displays various types of integrated optics beam
combiners for two telescopes. They can easily be upgraded to the
combination of a larger number of beams.  We have classified these beam
combiners with the same terminology as in Sect.\ \ref{sect:bc}. 

A co-axial beam combiner is made of waveguide junctions. Reverse Y-junctions
allow to collect only the constructive part of the interferometric signal
while X-crossing junctions with small angles get the whole interferometric
signal provided that asymmetric waveguides are used for the two arms. Note
that directional couplers can also be used despite the narrow bandpass.

A multi-axial beam combiner is formed by individual single mode waveguides
assembled by a taper that feed a planar waveguide. The light
propagates freely in the horizontal direction and the beams interfere at the
output of the device whereas light remains confined in the vertical
direction. The fringes can be sampled on a detector.

The multiplexer has no analogs in classical optics\footnote{except if we
pile up several coaxial beam combiners.}. The light from a given input beam
is mixed with the light from other input beams thanks to directional
couplers. The output beams are a linear combination of the input beams
whose ratios highly depend on the wavelength.

\subsection{Optical Path Difference modulators}

Small excursions are possible with integrated optics technologies. The
phase can be modulated up to 100 $\micron$ with on-the-chip electro-optics,
thermo-optics or magneto-optics actuators \citep{Alf82}. Such excursions
are long enough to modulate the optical path difference around the zero-OPD
location to scan the fringes.

\subsection{Wavelength selection}

Thin-film technology can be used to deposit any spectral filter at the
output of waveguides \citep{Ric96}. A particular application of the thin-film
coatings is the dichroic filters. Such components are usually integrated in
telecom devices and are attractive for astronomical interferometry in order
to perform various calibrations or controls.

\subsection{Photometric calibration}

Thanks to direct Y-junctions or direction couplers light can be partially 
extracted to achieve real-time photometric derivations.

\subsection{Polarization control}

The control of waveguide shapes permit to build polarizing components such as 
linear polarizers, polarization rotators or phase shifter \citep{Lang97}, 
which can be used to compensate residual instrumental polarization. 
In the future, integrated optics components could eventually be coupled 
with crystals (such as Lithium Niobate) which induces polarization thanks to 
Kerr or Pockels effects.

\subsection{Detection}

The size of waveguides ($1$ to $10\micron$) is similar to the size of
pixels in infrared arrays. Therefore, direct matching of the planar optics
component with an infrared detector would lead to a completely integrated
instrument with no relay optics between the beam combiner and the detector.
Furthermore recent developments of Supra-conducting Tunnel Junctions (STJ,
\citet{Fea98}) show that one may build pixel size detectors with photon
counting capabilities over a large spectral range (from ultra-violet to
near-infrared) with a very high quantum efficiency. Given its natural
spectral resolution (R=50) a STJ combined with an integrated interferometer
allows multichannel interferogram detection as well as fringe tracking
capabilities.  Since STJ are manufactured with the same etching technology
as some integrated optics component, one foresees a complete integrated
interferometer including one of the most sensitive detector.

In the future, detection techniques using parametric conversion
\citep{Rey96} could be implemented with optical waveguides.

\subsection{Switches}

Optical integrated switches \citep{Oll96} already exists and can be coupled
with an integrated interferometer to ensure the delay line function.

\section{Discussion}
\label{sect:disc}

Integrated optics is extremely attractive in astronomical interferometry for
combining two or more beams and for various functions \citep{Ker96}. 
In this section we discuss some intrinsic properties of integrated optics
components. This analysis leads us to list some specific advantages and
applications for this approach.

\subsection{Optical losses}

We have to distinguish several optical losses :
\begin{enumerate}
\item Fresnel losses.
  
  At the air/waveguide or air/coupling fiber interfaces, Fresnel losses
  occur. They equal about 4\% but can be reduced by anti-reflection coatings 
  deposited at the inputs and the outputs of the waveguides.
  
\item Coupling losses.  
  
  The light injection in the waveguide can be either direct or, more
  usually, via an optical fiber. According to the chosen solution, coupling
  losses exist at the air/waveguide interface or air/coupling fiber and
  at the fiber/waveguide interface.  For an efficient coupling,
  the incident energy has to match as much as possible
  with the propagating mode (numerical apertures, fiber core and waveguide 
  sizes, waveguide profile shape).
  
  All these conditions cannot be easily satisfied. In etching technology, the 
  process provides channels with non spherical sections, leading to
  coupling losses of about 0.33 dB or 7$\%$ excluding Fresnel losses). With 
  ion exchange technology, the coupling efficiency clearly depends on the 
  diffusion process and more specifically on the channel depth. The losses 
  are of the order of 2-3$\%$ if the waveguide is embedded inside the 
  substrate.

\item Propagation losses.
  
  Standard glasses provide low propagation losses for wavelengths less than
  $2.5 \micron$. With ion exchange technology, the propagation losses depend
  upon the diffused ions. For the more used ions ($\mbox{K}^+$,
  $\mbox{Tl}^+$ or $\mbox{Ag}^+$) these losses remain less than 0.2 dB/cm (a 
  1 cm-long component has a throughput of 94.5$\%$). Silicon etching 
  technology exhibits propagation losses of 5 dB/m. Therefore
  integrated optics cannot be used to realize lengthy components aimed at 
  transportation. 

\item Losses intrinsic to the integrated optics structure.
  
  Depending on the integrated optics design, light can be partially lost
  because of uncontrolled radiated modes like in the calssical reverse
  Y-junction. When used with two incident beams in opposite phases, the
  flux is radiated inside the substrate. This point is critical in
  astronomical interferometry where we wish to maximize the optical
  throughput. For this specific application, optimization and 
  simulation of various components are in progress \citep{Sch98}.

\end{enumerate}

\subsection{Spectral behavior}

\subsubsection{Available spectral ranges}

The off-the-shelves components are generally designed for telecom spectral
bands (0.8 $\micron$, 1.3 $\micron$ and 1.5 $\micron$). They can directly
be used to manufacture astronomical components for the I, J and H bands of
the atmosphere.  Standard glasses have an optical throughput higher than
90$\%$ in the visible and the near-infrared domain \cite[up to $2.5
\micron$, see][]{Sch96}. Ion exchange technology provide
integrated components for the K atmospheric bands ($2.2\micron$). For
higher wavelengths ($5 \micron$ or $10 \micron$), different technologies are 
under study.

Optical waveguides remain single-mode over a given spectral range (an 
octave in wavelength). This range is wide enough to cover a single
atmospheric band but not for several bands. However the compactness of
integrated optics components allow to use one optimized component for each
band without increasing the overall size of the instrument.

Y-junctions provides achromatic power division and beam combination which
make them attractive despite the loss of 50\% of the information in the
latter function. The other functions should be studied and optimized in order
to limit the chromatic dependence over the spectral range. Finally we
recommend to calibrate the device with spectral gain tables like
in standard astronomical imaging in order to suppress any device-dependent
effects.

\subsubsection{Chromatic dispersion}

Like fiber optics, integrated optics components have intrinsic chromatic
dispersion which could lead to a visibility loss over typical spectral
bandwidths of 0.2-0.4$\micron$ from the atmospheric bands. However the losses
are greatly reduced since:
\begin{itemize}
\item the mask has been designed to provide symmetrical interferometric
  arms with identical lengths, curvature, etc...;
\item the optical path difference between two arms is directly proportional
  to the length of the device.  For a typical length of a few centimeters,
  the optical length difference cannot exceed 100$\micron$ essentially due
  to machining defects (cutting and polishing);
\item the process for both technologies (ion exchange and etching) provides
  a good homogeneity for the index difference inside the waveguides.
\end{itemize}
Therefore even if we cannot preclude any contrast losses due to chromatic
dispersion, we think that this effect will remain small. 

However since the integrated optics component is part of an instrument,
special care must be taken to avoid other sources of chromatic
dispersion. In particular, optical fibers if used to inject stellar light
into the device have to be optimized accurately \citep{Rey96}.

\subsubsection{Dispersion capabilities}

Within the context of spectral interferometric measurements (Sect. 
~\ref{sect:spect}) the waveguide output is equivalent to the input slit of a 
spectrograph and is able to directly feed a spectrograph grating avoiding 
the cylindrical optics used to compress the Airy pattern in the direction
perpendicular to the fringes \citep{Pet98}.

\subsection{Polarization behaviour}
\label{sect:polar}

Both integrated optics technologies control the orientation of the neutral 
axes and thus provide components with intrinsic maintain of polarization 
properties. Provided that the design is symmetrical, the component does not 
introduce differential polarization, which is a crucial advantage for astronomical 
interferometry. Note that the coupling with polarization maintaining optical 
fibers has to be done with a great accuracy.  

\subsection{Thermal background}

Because of their small size, integrated optics components can easily be
integrated in a single camera dewar. Therefore no relay optics are needed
between the component and the detector, reducing the photon
losses. Moreover the waveguide can be cooled and put close to the detector
and the dewar can be blind, which reduces the thermal background.

\section{Conclusion and perspectives}

\begin{figure}[t]
  \begin{center}
    \leavevmode
    \includegraphics[angle=-90,width=0.7\columnwidth]{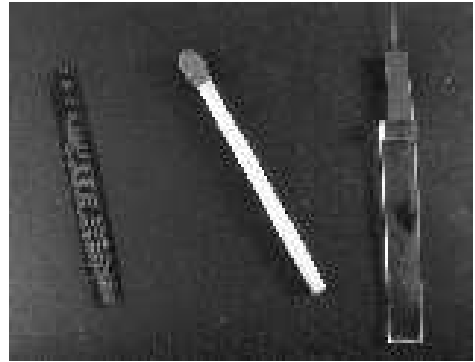}
    \caption{Photograph of two prototypes of integrated optics components
      together with a match to give the spatial scale. Left component
      features a 3-way beam combiner with photometric calibration channels
      manufactured with the silicon etching technology \citep{Sev99}. Right
      component manufactured with the ion exchange technology and connected
      with two fibers is a 2-way beam combiner with photometric calibration
      channels.}
    \label{fig:photo}
  \end{center}
\end{figure}

\subsection{Decisive advantages}

We have shown the great interest of using integrated optics for
astronomical interferometry. Figure \ref{fig:photo} shows a three-way beam
combiner with photometric calibrating channels made with the silicon
etching technology \citep{Sev99}.

We argue that integrated optics technology which is already industrially
mature,  presents the following main advantages for astronomical interferometry:
\begin{itemize}
\item Small size. A complete instrument can be integrated on a
  chip typically $5\mbox{ mm}\times20\mbox{ mm}$.
\item Stability. The instrument is completely stable while embedded in a
  substrate.
\item Low sensitivity to external constraints: temperature, pressure,
mechanical constrains.
\item Few opto-mechanical mounts and little alignment required. The only
  concern is coupling light into the waveguides.
\item Simplicity. For a complex instrument, the major efforts are shifted
  to the design of the mask, not in the construction phase.
\item Intrinsic polarization capabilities (Sect. ~\ref{sect:polar}).
\item Low cost. Integrated optics provides very low cost components
  and instrumentation set-up. Furthermore the price is the same for
  one or several components since the main cost is in the design and initial
  realization of the mask.
\end{itemize}

\subsection{Application to interferometry}
 
Integrated optics components are not well-suited for wide wavelength
coverage, high spectral dispersion and large field of view. Therefore, we
do not think that it will completely replace existing techniques in
astronomical interferometry. However with the characteristics presented in
this article, we think that integrated optics will be attractive for the
two following applications:

\begin{itemize}
\item Interferometers with a large number of apertures.\\
Whatever the complexity of the instrumental concept, only one component in
integrated optics allows combination of several beams and ensures the
photometric calibrations at low cost and with limited alignments

\item Space-based interferometers.\\
Integrated optics components with no internal alignments provide
reliable beam combiners for spatial interferometry.
\end{itemize}

These specific advantages lead us to perform laboratory experiments to
validate this approach. An interferometric workbench has been built to
completely characterize various components realized by both technologies.
First fringes with a white source have been obtained \citep[][, paper
II]{Ber99} and the integration of an interferometric instrument dedicated
to astronomical observations in the H and K atmospheric bands is in
progress \citep{Ber98}.

\section{Acknowledgments}

The authors are grateful to F.\ Reynaud (IRCOM - Univ.\ Limoges), P.\ 
Pouteau, P.\ Mottier and M.\ S\'everi (CEA/LETI - Grenoble) for fruitful
discussions, and to E.\ Le Coarer and P.\ Feautrier for the idea of
combining integrated optics and STJ. We would like to thank K.\ Wallace
for carefully reading the manuscript. The works have partially been funded
by PNHRA/INSU, CNRS/Ultimatech and DGA/DRET (Contract 971091).


\begin{thebibliography}{}
%
\bibitem[Alferness (1982)]{Alf82} 
  Alferness R.C. 1982, IEEE Transactions on Microwave Theory Techniques, 30,
  1121  
%
\bibitem[Baldwin et al.\ (1996)]{Bal96}
  Baldwin J.E., et al.\ 1996, A\&A 306, L13
%
\bibitem[Benson et al.\ (1997)]{Ben97}
  Benson J.A., et al.\ 1997, AJ 114, 1221
%
\bibitem[Berger et al.\ (1998)]{Ber98}
  Berger J.-P.,  Rousselet-Perraut K., Kern P., Malbet F., Schanen-Duport I., 
  Nabias L., Benech P.\ 1998, Integrated Optics components for 
  interferometric beam combination.  In: Reasenberg R.D.\ (ed.) Proc.\ SPIE 
  3350, Astronomical Interferometry, p.\ 898.
%
\bibitem[Berger et al.\ (1999)]{Ber99}
  Berger J.-P.,  Rousselet-Perraut K., Kern P., Malbet F., Schanen-Duport I., 
  Reynaud F., Haguenauer P., Benech P.\ 1999, A\&A, in press (paper II).
%
\bibitem[Carleton et al.\ (1994)]{Car94} 
  Carleton N. P., et al.\ 1994, Current status of the IOTA
  interferometer. In: Breckinridge J.B.\ (ed.)  Proc.\ SPIE 2200, Amplitude
  and Intensity Spatial Interferometry II, 152. 
%
\bibitem[Colavita et al.\ (1991)]{Col91} 
  Colavita M.M., Hines B.E., Shao M.\ Klose G.J., Gibson B.V.\ 1991,
  Prototype high speed optical delay line for stellar interferometry. In:
  Ealey M.A.\ (ed.)  Proc.\ SPIE 1542, Active and Adaptive Optical
  Systems, 205 
%
\bibitem[Colavita et al.\ (1994)]{Col94} 
  Colavita M.M., et al.\ 1994, ASEPS-0 testbed interferometer. In:
  Breckinridge J.B.\ (ed.)  Proc.\ SPIE 2220, Amplitude and Intensity Spatial
  Interferometry II, 89
%
\bibitem[Colavita et al.\ (1998)]{Col98}
  Colavita M.M., et al.\ 1998, Keck Interferometer. In: Reasenberg R.D.,
  Unwin S.C.\ (eds.) Proc.\ SPIE 3350, Astronomical Telescopes and
  Instrumentation: Astronomical Interferometry, in press
%
\bibitem[Connes et al.\ (1984)]{Con84} 
  Connes P., Froehly C., Facq P.\ 1984, A Fiber-Linked Version of Project
  TRIO. In: Longdon N., Melita O.\ (eds.)  Proc.\ ESA Colloq., Kilometric
  Optical Arrays in Space.\ ESA, Carg\`ese, p.\ 49
%
\bibitem[Coud\'e du Foresto (1996)]{For96a} 
  Coud\'e du Foresto V.\ 1996, Fringe Benefits: the Spatial Filtering
  Advantages of Single-Mode Fibers. In: Kern P., Malbet F.\  (eds) Proc.
  AstroFib'96, Integrated Optics for Astronomical Interferometry.
  Bastianelli-Guirimand, Grenoble, p.\ 27
%
\bibitem[Coud\'e du Foresto \& Ridgway (1991)]{For91} 
  Coud\'e du Foresto V., Ridgway S.\ 1991, FLUOR: a Stellar Interferometer
  Using Single-Mode Infrared Fibers. In: Beckers J., Merkle F.\ (eds.) Proc.
  ESO conf., High-resolution imaging by interferometry II.\ ESO, Garching,
  731 
%
\bibitem[Coud\'e du Foresto et al.\ (1996)]{For96b} 
  Coud\'e du Foresto V., Perrin G., Mariotti J.-M., Lacasse M., Traub W.
  1996, The FLUOR/IOTA Fiber Stellar Interferometer. In: Kern P., Malbet F.
  (eds) Proc.\  AstroFib'96, Integrated Optics for Astronomical
  Interferometry.\  Bastianelli-Guirimand, Grenoble, p.\ 115
%
\bibitem[Davis et al.\ (1994)]{Dav94}
  Davis J., Tango W.J., Booth A.J., Minard R.A., Owens S.M., Shobbrook R.R.
  1985, Progress in commissioning the Sydney University Stellar
  Interferometer (SUSI). In: Breckinridge J.B.\ (ed.)  Proc.\ SPIE 2220,
  Amplitude and Intensity Spatial Interferometry II, 231
%
\bibitem[Fizeau (1868)]{Fiz68}
  Fizeau H.\ 1868,  C.\ R.\ Acad.\ Sci.\ Paris, 66, 932
%
\bibitem[Froehly (1981)]{Fro81} 
  Froehly C.\ 1981, Coherence and Interferometry through Optical Fibers. In:
  Ulrich M.H., Kj\"ar K.\ (eds.)  Proc.\ ESO conf., Science Importance of
  High Angular Resolution at Infrared and Optical Wavelengths.\ ESO,
  Garching, p.\ 285
%
\bibitem[Jeunhomme (1990)]{Jeu90} 
  Jeunhomme L.\ 1990, Single-mode fiber optics, Marcel Dekker Inc.
%
\bibitem[Kern et al.\ (1996)]{Ker96} 
  Kern P., Malbet F., Schanen-Duport I., Benech P.\ 1996, Integrated optics 
  single-mode interferometric beam combiner for near infrared astronomy. In: 
  Kern P., Malbet F.\ (eds) Proc.\  AstroFib'96, Integrated Optics for 
  Astronomical Interferometry.\  Bastianelli-Guirimand, Grenoble, p.\ 195
%
\bibitem[Feautrier et al.\ (1998)]{Fea98}
  Feautrier P., et al.\ 1998, Superconducting Tunnel Junctions for photon 
  counting in the near infrared wavelengths. In: Applied Superconductivity 
  Conference, Palm Desert
%
\bibitem[Koechlin et al.\ (1996)]{Koe96}
  Koechlin L., et al.\ 1996, Appl.\ Opt.\ 35, 3002
%
\bibitem[Labeyrie (1975)]{Lab75}
  Labeyrie A.\ 1975, ApJ 196, L71-L75
%
\bibitem[Lang (1997)]{Lang97}
  Lang T.\ 1997, PhD Thesis, University of Grenoble, France.
%
\bibitem[Lef\`evre (1980)]{Lef80}
  Lef\`evre H.C.\ 1980, Electron.\ Letters 16,778
%
\bibitem[Roddier \& L\'ena (1984)]{Len84a}
  Roddier F., L\'ena P.\ 1984, J.\ Opt.\ 15, 171
%
\bibitem[Roddier \& L\'ena (1984)]{Len84b}
  Roddier F., L\'ena P.\ 1984, J.\ Opt.\ 15, 363
%
\bibitem[Mariotti (1998)]{Mar98}
  Mariotti J.-M.\ 1998, VLTI: a Status Report. In: Reasenberg R.D., Unwin
  S.C.\ (eds.) Proc.\ SPIE 3350, Astronomical Telescopes and Instrumentation:
  Astronomical Interferometry, in press
%
\bibitem[Mariotti \& Ridgway (1988)]{MRi88}
  Mariotti J.-M., Ridgway S.\ 1988, A\&A 195, 350
%
\bibitem[Mariotti et al.\ (1992)]{Mar92}
  Mariotti J.-M., et al.\ 1992, Coherent Combined Instrumentation for the
  VLT Interferometer.\ VLT Report No 65, ESO, Garching
%
\bibitem[Mariotti et al.\ (1996)]{Mar96}
  Mariotti J.-M., Coud\'e du Foresto V., Perrin G., Zhao P., L\'ena P.
  1996, A\&AS 116, 381
%
\bibitem[McAlister et al.\ (1994)]{McA94}
  McAlister H.A., et al.\ 1994, CHARA Array. In: Breckinridge J.B.\ (ed.)
  Proc.\ SPIE 2200, Amplitude and Intensity Spatial Interferometry II, 129.
%
\bibitem[Michelson \& Pease (1921)]{Mic21}
  Michelson A.A., Pease F.G.\ 1921, ApJ 53, 249
%
\bibitem[Miller (1969)]{Mil69}
  Miller S.E.\ 1969, The Bell System Technical Journal, vol.\ 48, 2059
%
\bibitem[Mottier (1996)]{Mot96} 
  Mottier P.\ 1996, Integrated Optics and Micro-Optics at LETI. In: Kern P.,
  Malbet F.\ (eds) Proc.\  AstroFib'96, Integrated Optics for Astronomical
  Interferometry.\  Bastianelli-Gui\-ri\-mand, Grenoble, p.\ 63
%
\bibitem[Mourard et al.\ (1994)]{Mou94}
  Mourard D., Tallon-Bosc I., Blazit A., Bonneau D., Merlin G., Morand F.,
  Vakili F., Labeyrie A.\ 1994, A\&A 283, 705
%
\bibitem[Ollier \& Mottier (1996)]{Oll96} 
  Ollier E. \& Mottier P.\ 1996, Electronics Letters, Vol.\ 32, 21
%
%
\bibitem[Petrov et al.\ (1998)]{Pet98} 
  Petrov R.G., Malbet F., Richichi A., Hofmann K.-H. 1998,
  The ESO Messenger, 92, 11
\bibitem[Rabbia et al.\ (1996)]{Rab96} 
  Rabbia Y., M\'enardi S., Reynaud F., Delage L.\ 1996, The ESO-VLTI fringe
  sensor. In: Kern P., Malbet F.\ (eds) Proc.\  AstroFib'96, Integrated Optics
  for Astronomical Interferometry.\  Bastianelli-Guirimand, Grenoble, p.\ 175
%
\bibitem[Ramaswamy \& Srivastava (1988)]{Ram88}
  Ramaswamy R.V., Srivastava R.\ 1988, J.\ of Light.\ Tech.\ 6, 984
%
\bibitem[Reynaud (1993)]{Rey93} 
  Reynaud F.\ 1993, Pure Applied Optics 2, 185-188.
%
\bibitem[Reynaud \& Delaire (1994)]{ReD94} 
  Reynaud F., Delaire E.\ 1994,
  Linear Optical Path Modulation with a Lambda/200 Accuracy Using a Fiber
  Stretcher. In: Cerutti-Maori M.-G., Roussel Ph.\ (eds.), Proc.\ SPIE 2209,
  Space Optics 1994: Earth Observation and Astronomy, 431
%
\bibitem[Reynaud \& Lagorceix (1996)]{Rey96} 
  Reynaud F, Lagorceix H.\ 1996, Stabilization and Control of a Fiber Array
  for the Coherent Transport of Beams in a Stellar Interferometer. In: Kern
  P., Malbet F.\ (eds) Proc.\  AstroFib'96, Integrated Optics for
  Astronomical Interferometry.\  Bastianelli-Guirimand, Grenoble, p.\ 249
%
\bibitem[Reynaud et al.\ (1994)]{Rey94}
  Reynaud F, Alleman J.J., Lagorceix H.\ 1994, Interferometric fiber arms
  for stellar interferometry. In: Cerutti-Maori M.-G., Roussel Ph.\ (eds.),
  Proc.\ SPIE 2209, Space Optics 1994: Earth Observation and
  Astronomy, 431
%
\bibitem[Richier (1996)]{Ric96} 
  Richier R.\ 1996, End-Coating in Optical Fibers Developments and
  Application. In: Kern P., Malbet F.\ (eds) Proc.\  AstroFib'96, Integrated
  Optics for Astronomical Interferometry.\  Bastianelli-Guirimand, Grenoble,
  p.\ 163
%
\bibitem[Ross (1989)]{Ros89}
  Ross L.\ 1989, Glastechniche Berichte 62, 285
%
\bibitem[Rousselet-Perraut et al.\ (1996)]{Rou96}
  Rousselet-Perraut K., Vakili F., Mourard D.\ 1996, Optical Engineering
  Vol.\ 35, 10, 2943
%
\bibitem[Rousselet-Perraut et al.\ (1998)]{Rou98}
  Rousselet-Perraut K.,Hill L., Lasselin-Waultier G., Boit J.L., Rousset G., Blanc J.C.\ and Voet C.\ 1998, Optical Engineering
  Vol.\ 37, 2, 610
%
\bibitem[Schanen-Duport et al.\ (1996)]{Sch96} 
  Schanen-Duport I., Benech P., Kern P., Malbet F.\ 1996, Optical Waveguides
  Made by Ion Exchange for Astronomical Interferometry Applications at the
  Wavelength of 2.2 Microns. In: Kern P., Malbet F.\ (eds) Proc.
  AstroFib'96, Integrated Optics for Astronomical Interferometry.
  Bastianelli-Guirimand, Grenoble, p.\ 99
%
\bibitem[Schanen-Duport et al.\ (1998)]{Sch98} 
  Schanen-Duport I., El-Sabban S., Berger J.P., Kern P., Malbet F., Rousselet-
  Perraut K.\ 1998, Proc.\ JNOG, Marly-le-Roi.
%
\bibitem[Severi et al.\ (1999)]{Sev99}
  Severi M., Pouteau P., Mottier P., Kern P. 1999, A waveguide
  interferometer for phase closure in astronomy. In: ECIO'99, Torino, 15
  April 1999. 
%
\bibitem[Shaklan (1990)]{Sha90}
  Shaklan S.B.\ 1990, Optical Engineering 29, 684
%
\bibitem[Shaklan \& Roddier (1987)]{Sha87}
  Shaklan S.B., Roddier F.\ 1987, Applied Optics 26, 2159
%
\bibitem[Shaklan \& Roddier (1988)]{Sha88}
  Shaklan S.B., Roddier F.\ 1988, Applied Optics 27, 2334
%
\bibitem[Shao \& Staelin (1977)]{Shao77} 
  Shao M., Staelin D.H.\ 1977, JOSA 67, 81
%
\bibitem[Shao et al.\ (1988)]{Shao88}
  Shao M., Colavita M.M., Hines B.E., Staelin D.H., Hutter D.J.\ 1988, A\&A
  193, 357
%
\bibitem[Simohamed \& Reynaud (1996)]{Sim96} 
  Simohamed L.M., Reynaud F.\ 1996, A Two Meter Stroke Optical Fibre Delay
  Line. In: Kern P., Malbet F.\ (eds) Proc.\  AstroFib'96, Integrated Optics
  for Astronomical Interferometry.\  Bastianelli-Guirimand, Grenoble, p.\ 217
%
\bibitem[Stefan (1874)]{Ste74}
  St\'ephan E.\ 1874,  C.\ R.\ Acad.\ Sci.\ Paris, 78, 1008
%
\bibitem[Strohh\"ofer et al. (1998)]{Stro98}
  Strohh\"ofer et al.\ 1998, Active optical properties of Erbium-doped 
  Ge$O_2$-based sol-gel planar waveguides, In: Thin Solid Films, in press  
%
\bibitem[Turner \& Brummelaar (1997)]{Tur97}
  Turner N.H., Brummelaar T.\ 1997, BAAS 190
%
\bibitem[White et al.\ (1994)]{Whi94}
  White N.M., et al.\ 1994, Progress Report on the Construction of the Navy
  Prototype Optical Interferometer at the Lowell Observatory. In:
  Breckinridge J.B.\ (ed.)  Proc.\ SPIE 2220, Amplitude and Intensity Spatial
  Interferometry II, 242.
%
\bibitem[Zernike \& Midwinter (1973)]{Zer73}
Zernike F., Midwinter J.E.\ 1973, Applied nonlinear optics, Wiley sons

%
\bibitem[Zhao et al.\ (1995)]{Zha95}
  Zhao P., Mariotti J.-M., Coud\'e du Foresto V., L\'ena P., Perrin
  G.\ 1995, Multistage Fiber Optic Delay Line for Astronomical
  Interferometry. In: Barden S.C.\ (ed.), proc.\ SPIE 2476, Fiber Optics in
  Astronomical Applications, 108
%
\end{thebibliography}
\end{document}